\documentclass[aps,prl,reprint,twocolumn,superscriptaddress,amsmath,amssymb,citeautoscript]{revtex4-1}
\usepackage{amsmath}
\usepackage{amssymb}
\usepackage{graphicx}
\usepackage{color}
\usepackage{bm}
\usepackage{float}

\usepackage{epstopdf}
\usepackage{grffile}
\DeclareGraphicsExtensions{.eps}

\newcommand{\e}{{\rm e}}

\newcommand{\ii}{{\rm i}}

\newcommand{\la}{{\langle}}
\newcommand{\ra}{{\rangle}}
\newcommand{\PO}{\mathcal{O}_{P}^2}
\newcommand{\SO}{\mathcal{O}_{S}^2}

\begin{document}

\title{Hidden order and symmetry protected topological states in quantum link ladders}

\author{L. Cardarelli}
\author{S. Greschner}
\author{L. Santos}
\affiliation{Institut f\"ur Theoretische Physik, Leibniz Universit\"at Hannover, 30167~Hannover, Germany}

%%%%%%%%%%%%%%%%%%%%%%%%%%%%%%%%%%%%%%%%%%%%%%%%%%%%%%%%%%%%%%%%%%%%%%%%%%%%%%%%%%%%%%%%

\begin{abstract}
We show that whereas spin-$1/2$ one-dimensional U(1) quantum-link models~(QLMs) are topologically trivial, when implemented in ladder-like lattices these models may present an intriguing ground-state phase diagram, 
which includes  a symmetry protected topological~(SPT) phase that may be readily revealed by analyzing long-range string spin correlations along the ladder legs.
We propose a simple scheme for the realization of spin-$1/2$ U(1) QLMs based on single-component fermions loaded in an optical lattice with $s$- and $p$-bands, 
showing that the SPT phase may be experimentally realized by adiabatic preparation.
\end{abstract}

%%%%%%%%%%%%%%%%%%%%%%%%%%%%%%%%%%%%%%%%%%%%%%%%%%%%%%%%%%%%%%%%%%%%%%%%%%%%%%%%%%%%%%%%

\date{\today}

\maketitle

%%%%%%%%%%%%%%%%%%%%%%%%%%%%%%%%%%%%%%%%%%%%%%%%%%%%%%%%%%%%%%%%%%%%%%%%%%%%%%%%%%%%%%%%

% INTRODUCTION

The realization of lattice gauge models using ultra cold gases has attracted a major theoretical attention in recent years~\cite{Wiese2013,Wiese2014,Zohar2016,Dalmonte2016}. 
Various ideas for creating dynamical gauge fields have been proposed~\cite{Buchler2005,Cirac2010,Weimer2010,Zohar2011,Kapit2011,Zohar2012,Banerjee2012,Zohar2013,Banerjee2013,Zohar2013b,Tagliacozzo2013,Hauke2013,Kasper2017}. 
Recently the Schwinger model has been simulated in ion chains~\cite{Martinez2016}. Particular interest has been devoted to 
quantum-link models~(QLMs)~\cite{Chandrasekharan1997}, which generalize lattice gauge theory~\cite{Wilson1974} by realizing
continuous gauge symmetries with discrete gauge variables~(quantum links).  QLMs  are relevant in particle physics, and in particular QCD~\cite{Brower1999}, and  
in condensed matter physics~\cite{Hermele2004,Levin2005}. In U(1) QLMs, links are represented by quantum spins and fermions provide 
the matter field, making these QLMs particularly suitable for simulation with cold lattice gases.

In this Letter we study the topological properties of spin-$1/2$ U(1) QLMs.  Topological quantum systems have become one of the most active research areas 
during the past decades~\cite{Hasan2010,XiaoLiang2011}. In particular the understanding of topological phases in strongly correlated quantum systems remains challenging. 
The study of symmetry protected topological~(SPT) states has triggered a large progress in this field~\cite{Senthil2015}. 
SPT phases have been classified by means of entanglement properties and group theoretical 
considerations~\cite{Pollmann2010, Schuch2011, Fidkowski2011, Pollmann2012, Pollmann2012A, Chen2012}. 
Indeed in one-dimensional~(1D) systems, SPT phases are the only realizable class of topological quantum states, a prominent example being the so-called Haldane phase of 
odd-integer spin chains~\cite{Affleck1987,Kiall2013}. Generalizations of the Haldane phase have been theoretically studied in the context of ultra-cold 
gases~\cite{DallaTorre2006, Berg2008, Greschner2013, Nonne2010, Jaramillo2013, Lange2017}.

Real or synthetic ladder-like lattices have recently constituted the focus of major efforts~\cite{Atala2014,Mancini2015, Stuhl2015} in the context of the realization of static gauge fields in ultra-cold atomic systems.
We show below that although in 1D spin-$1/2$ U(1) QLMs are topologically trivial, when implemented in ladder-like lattices these models present an intriguing ground-state phase diagram, 
which interestingly includes an SPT phase that we characterize using a generalized topological order parameter and the entanglement spectrum. 
We show that the SPT phase may be revealed by analyzing string spin correlations along the ladder legs.
Moreover, we propose a simple scheme for the realization of the QLM based on $s$-$p$ lattices~\cite{Wirth2011}, showing that the SPT phase may be experimentally 
realized by adiabatic preparation.

%%%%%%%%%%%%%%%%%%%%%%%%%%%%

% FIGURE 1

\begin{figure}[bt]
\centering
\includegraphics[width=1\linewidth]{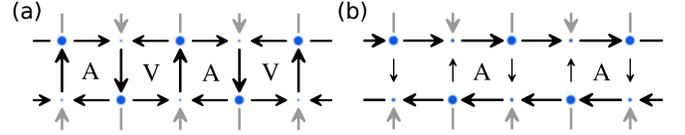}\\
\caption{Ground state of the QLL for $\mu=-|J_x|$ and (a) $J_y/J_x=0.2$~(VA phase) and (b) $1.8$~(V0 phase). The length and size of the arrows and the size of the points is 
proportional to $\langle \tilde S^z_{i,j;i+1,j} \rangle$ and $\langle \Psi_{i,j}^\dag \Psi_{i,j}\rangle$, respectively. For bonds along the legs we denote for 
convenience $\uparrow$~($\downarrow$) as $\rightarrow$~($\leftarrow$). 
V~(A) denote (anti)vortex-like spin configurations. In the SPT phase~(not shown) local spin-expectations vanish and the fermions are evenly distributed. 
Grey arrows indicate boundary conditions (see text).}
\label{fig:patterns}
\end{figure}

%%%%%%%%%%%%%%%%%%%%%%%%%%%%

%%%%%%%%%%%%%%%%%%%%%%%%%%%%

% FIGURE 2

\begin{figure*}[t]
\centering
\includegraphics[width=0.32\linewidth]{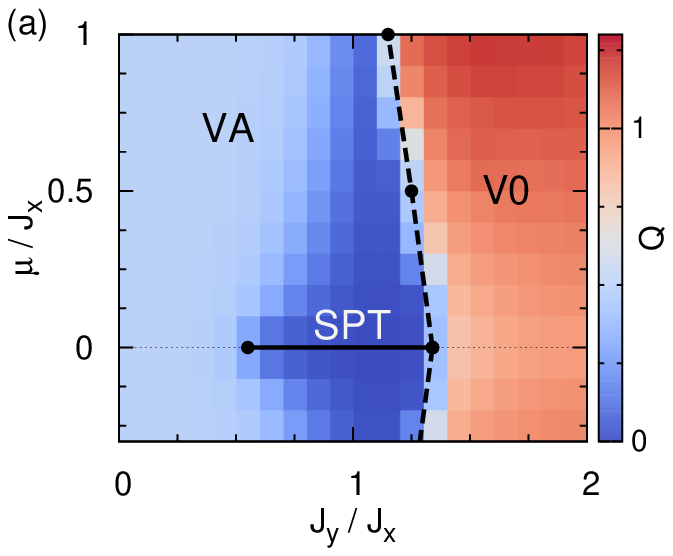}
\includegraphics[width=0.32\linewidth]{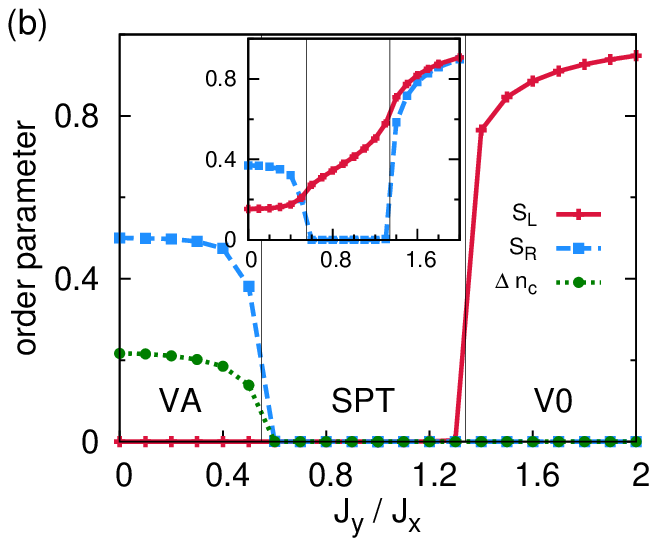}
\includegraphics[width=0.32\linewidth]{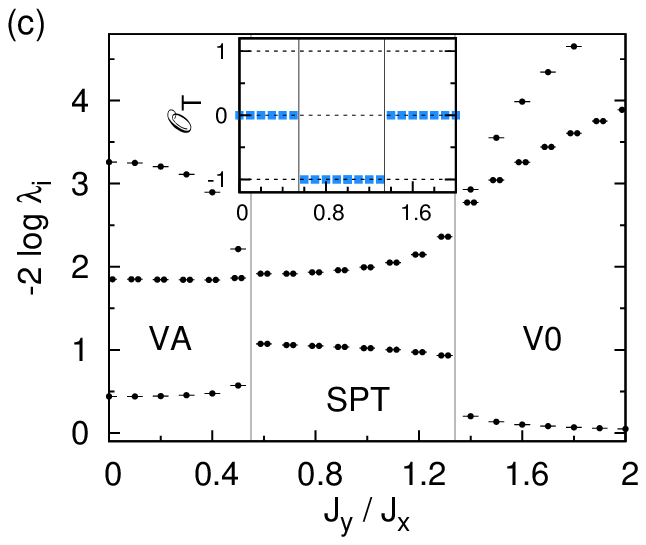}
\caption{(a) Phase diagram for the QLL as a function of $\mu/J_x$ and $J_y/J_x$ obtained from infinite time evolving block decimation (iTEBD) simulations~\cite{Vidal2007} with up to $80$ matrix states. 
The mirror symmetry of the ladder results in the same phase diagram for $\mu>0$ and $\mu<0$.
Only for $\mu=0$ the SPT phase is realized. The phase transition points (indicated by solid circles) are determined by means of density matrix renormalization group (DMRG)~\cite{White1992, Schollwoeck2011} 
simulations with up to $800$ states and open boundary conditions~\cite{SMnote}. (b) $S_L$, $S_R$ and $\Delta n_c$ for $\mu=0$. 
The inset shows $\PO$~(blue dashed line) and $\SO$~(red solid line) for the same parameters.
(c) Largest values of the entanglement spectrum $\lambda_i$ for $\mu=0$. In the SPT phase the spectrum is doubly degenerate. The inset shows the 
generalized topological order parameter $\mathcal{O}_T$.}
\label{fig:pd_cuts_m0}
\end{figure*}

%%%%%%%%%%%%%%%%%%%%%%%%%%%%

%%%%%%%%%%%%%%%%%%%%%%%%%%%%%%%%%%%%%%%%%%%%%%%%%%%%%%%%%%%%%%%%%%%%%%%%%%%%%%%%%%%%%%%%

% MODEL

\paragraph{Model.--} We introduce a two-legs-ladder extension of the QLM, which we call {\em quantum link ladder} (QLL): 
\begin{eqnarray}
H_{\mathrm {QLL}} & = & \mu \sum_{i,j} (-1)^{i+j} n_{i,j} \nonumber \\
&-& J_x  \sum_{i,j} \left ( \Psi_{i,j}^\dagger \tilde S^+_{i,j; i+1,j} \Psi_{i+1,j} + {\mathrm {H. c.}} \! \right ) \nonumber \\
&- &J_y  \sum_{i} \left ( \Psi_{i,1}^\dagger \tilde S^+_{i,0; i,1} \Psi_{i,0} + {\mathrm {H. c.}} \! \right ),
\label{eq:QLL}
\end{eqnarray}
where $\Psi_{i,j}$ are staggered fermionic operators at rung $i$ of leg $j=0$~(upper) and $1$~(lower), and $J_x$~($J_y$) is the hopping along the legs~(rungs). 
We define the A~(B) sites as those with even~(odd) $i+j$, which have on-site energy $\mu$~($-\mu$). 
In the analogy with QCD~\cite{Wiese2013}, filled A sites correspond to particles and empty B sites to anti-particles, with $\mu$ acting as particle mass.  
The gauge field characterizing the bond between nearest-neighboring sites, is represented for the Abelian case by a spin-$S$ operator~\cite{Wiese2013}. 
We assume $S=1/2$, and hence the gauge field is given by spin-$1/2$ operators $\tilde S^\pm$. 

%%%%%%%%%%%%%%%%%%%%%%%%%%%%%%%%%%%%%%%%%%%%%%%%%%%%%%%%%%%%%%%%%%%%%%%%%%%%%%%%%%%%%%%%

% 1D QLM

\paragraph{1D QLM.--} 
We evaluate first the simplest and best understood case of a 1D QLM, which results by considering a single leg~($j=0$) in Eq.~\eqref{eq:QLL}.
We consider only states that obey a local gauge symmetry~(Gauss' law):
 $\tilde S_{i,i+1}^z - \tilde S_{i-1,i}^z = n_i -\epsilon_i$, 
with $\epsilon_{i\in A}=0$ and $\epsilon_{i\in B}=1$ .  

In the large mass limit, $|\mu|\gg J_{x}$ we integrate out the particle motion, working for 
$\mu\to -\infty$ in the manifold in which the $A$~($B$) sublattice is fully occupied~(empty). In the 1D QLM, the ground-state is uniquely determined by Gauss's law, being 
a zero net flux (Z) phase~\cite{Rico2014}, in which filled $A$~(empty $B$) sites are accompanied by outgoing~(incoming) spins, $|0\rangle_A\equiv |\leftarrow 1 \rightarrow \rangle$ and 
$|0\rangle_B\equiv |\rightarrow 0 \leftarrow \rangle$, where we employ the spin notation introduced in Fig.~\ref{fig:patterns}.

For finite $\mu/J_x$,  the Z phase presents defects:  
$|+'\rangle_B=|\leftarrow 1 \leftarrow\rangle$, $|-'\rangle_B=|\rightarrow 1 \rightarrow\rangle$, 
$|-'\rangle_A=|\leftarrow 0 \leftarrow\rangle$ and $|+'\rangle_A=|\rightarrow 0 \rightarrow\rangle$. 
We define the magnetization for a site $i$ as $S_i^z=(-1)^{i} (\tilde S^z_{i-1,i} +  \tilde S^z_{i,i+1})$, and 
evaluate the parity order 
$\mathcal{O}_{P}^2 = \lim_{(k-j)\to \infty}\e^{\ii \pi \sum_{k<l<j} S_{i}^z}$, and string order 
$\mathcal{O}_{S}^2 = \lim_{(k-j)\to \infty} S_k^z \e^{\ii \pi \sum_{k<l<j} S_l^z} S_i^z$.
Gauss' law breaks the $\mathbb{Z}_2$ chiral symmetry~\cite{Rico2014}, and hence 
the defects on top of the Z phase are directed, i.e. they are strictly formed in $|-' , +'\rangle_{i,i+1}$ pairs. Moreover, a defect pair cannot split 
due to Gauss' law.  This selective pair creation induces $\PO\neq 0$ and $\SO\neq 0$ for any $\mu$~\cite{Greschner2014}, precluding 
a Haldane-like phase~(which would have  $\PO=0$ and $\SO\neq 0$). At
$\mu/J_x\simeq 0.45$ there is an Ising-like phase transition into the so-called non-zero flux~(NZ) phase~\cite{Rico2014}. This phase, which for $\mu\to\infty$ is a N\'eel-like state of $|\pm'\rangle$ defects, 
also presents $\PO,\SO \neq 0$~\cite{SMnote}.

%%%%%%%%%%%%%%%%%%%%%%%%%%%%%%%%%%%%%%%%%%%%%%%%%%%%%%%%%%%%%%%%%%%%%%%%%%%%%%%%%%%%%%%%

% QLL

\paragraph{QLL.--}  As for the 1D QLM, in the QLL we are only interested in states that obey Gauss' law: 
\begin{equation}
\tilde S^z_{i,j;i+1,j}-\tilde S^z_{i-1,j;i,j}+\tilde S^z_{i,j;i,j-1}-\tilde S^z_{i,j;i,j-1}=n_{i,j}-\epsilon_{i,j}, \!\!
\end{equation}
with $\epsilon_{i,j \in A}=0$ and $\epsilon_{i,j \in B}=1$.  Note that the orientation of the virtual spins placed outside the ladder~(in grey in Figs.~\ref{fig:patterns}) remains fixed, 
resulting in boundary conditions for the possibly QLL states. As shown below, fixing by construction staggered boundary conditions~(see Figs.~\ref{fig:patterns}) 
results in an intriguing physics for the QLL. We consider below $\mu<0$, but, contrary to 1D, the spatial mirror symmetry of the ladder-like lattice results in an identical ground-state phase diagram for $\mu>0$.

Whereas in the 1D QLM Gauss' law fixes a unique ground-state for large $|\mu|$, this is not the case in the QLL. For $\mu\to -\infty$, 
$A$ sites and their neighboring spins may be in three states: 
$|0\rangle_A \equiv |\!\leftarrow 1 \rightarrow \rangle$, $|+\rangle_A \equiv |\!\leftarrow 1 \leftarrow \rangle$, and $|-\rangle_A \equiv |\!\!\rightarrow 1 \rightarrow \rangle$. Note that due to 
Gauss' law and the boundary conditions the orientation of the spin on the rung is determined once the left and right spins are chosen. Similarly for  $B$ sites only three states are possible:
$|0\rangle_B \equiv |\!\rightarrow 0 \leftarrow \rangle$, $|+\rangle_B \equiv |\!\rightarrow 0 \rightarrow \rangle$, and $|-\rangle_B \equiv |\!\leftarrow 0 \leftarrow \rangle$.
For a given rung, irrespective of whether the upper site is A  or  B, only three states are relevant~\cite{otherstatesnote}: 
$|\phi_0\rangle = \left | \begin{array}{c} 0  \\   0   \end{array} \right \rangle$, 
$|\phi_+\rangle = \left | \begin{array}{c} +  \\ +   \end{array} \right \rangle$, 
$|\phi_-\rangle = \left | \begin{array}{c} -  \\ -   \end{array} \right \rangle$.  
These rung states form an effective spin-$1$ system, which up to order $J_{x,y}^4/\mu^3$ is determined by the Hamiltonian:
\begin{equation}
 H_{LM} \!=\! \sum_i \!\left [ D(R_i^z)^2 \! +\! K \! \left ( \left ( R_i^+ R_{i+1}^-\right )\!\! \left (R_i^z  R_{i+1}^z \right )\!\!+\! \mathrm{H. c.}\! \right ) \right ]\!, \!\!\!
 \label{eq:LM}
\end{equation}
where we define in rung $i$ the spin-$1$ operators $R_i^{\pm,z}$ in the basis $\{|\phi_0\rangle, |\phi_\pm\rangle\}$, $D=(J_x^2-J_y^2)/2\hbar^2|\mu|$ 
, and $K=J_y^2J_x^2/8\hbar^4|\mu|^3$ results from ring-exchange processes. 

For sufficiently large $D/K>0$, i.e. $J_y/J_x<1$, the 
phase in which all rungs are in $|\phi_0\rangle$ is favored. This is similar to the large-$D$ phase of spin-$1$ systems~\cite{Chen2003}, or the Mott phase in Hubbard models.
This phase corresponds to the vortex-antivortex~(VA) configuration depicted in Fig.~\ref{fig:patterns}a.
On the contrary for large $D/K<0$, i.e. $J_y/J_x>1$, a double-degenerate N\'eel-like phase $|\dots \phi_+,\phi_-,\phi_+,\phi_- \dots\rangle$ is the ground-state, which is analogous to the 
density-wave phase found in extended Hubbard models.  This phase corresponds to the configuration of Fig.~\ref{fig:patterns}b, characterized by 
vortices separated by a plaquette without vorticity~(V0 phase). 

Crucially the ring-exchange does not lead to 
a regular XY spin-exchange in Eq.~\eqref{eq:LM}, since due to Gauss' law only processes $|\phi_0,\phi_0\rangle_{i,i+1} \leftrightarrow |\phi_-,\phi_+ \rangle_{i,i+1}$ are allowed. 
As a result, whereas in the vicinity of $D\simeq 0$ a Haldane phase is expected for the spin-$1$ XY model with single-ion anisotropy~\cite{Chen2003}, 
we just observe for large $\mu$ a single phase transition between the VA and the V0 phase which is second-order due to the finite ring-exchange.

As for the 1D QLM we introduce for each leg the site magnetization $S_{i,j}^z=(-1)^{i+j} (\tilde S^z_{i-1,j;i,j} +  \tilde S^z_{i,j;i+1,j})$.  
Figure~\ref{fig:pd_cuts_m0}~(a) depicts $Q=S_L+S_R$, where $S_L=\frac{1}{L}\sum_i \la |S^z_{i,j}| \ra$ characterizes the leg spins, 
$S_R=\frac{1}{L}\sum_i \langle |\tilde S^z_{i,0;i,1}| \rangle$ characterizes the rung spins, and $L$ is the number of rungs. 
Note that $Q=1/2$ in the defect-free VA phase, whereas $Q=3/2$ in the defect-free V0 phase. Hence the VA-V0 transition~(dashed line) in the large mass limit is characterized by 
an abrupt jump in the value of $Q$. For finite $\mu$, $Q$ significantly decreases within the VA phase when approaching the phase transition (deep blue region). 
This decrease is connected to the appearance of defects in the VA phase ($|0'\rangle_A\equiv |\!\! \rightarrow 0 \leftarrow\!\rangle_A$, $|0'\rangle_B\equiv |\!\leftarrow 1 \rightarrow\!\rangle_B$, 
$|\pm'\rangle_A$, $|\pm'\rangle_B$). Although the reduction of $Q$ does not result into a phase transition~\cite{SMnote}, the crossover within the VA phase is evident and heralds the appearance of the SPT phase discussed below.

From the site magnetization $S_{i,j}^z$ we evaluate the corresponding $\PO$ and $\SO$ along the upper~(or lower) leg.
Gauss' law induces $\SO\neq 0$ for any $\mu$ and $J_y/J_x$. However, contrary to 1D, the ladder geometry permits the breaking of defect pairs along the leg created on top of the VA phase, 
and hence $\PO$ may in principle vanish. We observe however $\PO\neq 0$ for any $|\mu|>0$~\cite{SMnote}, in accordance with the observation that the large-$\mu$ phases VA and V0 evolve 
adiabatically without crossing any phase transition down to zero mass. 

%%%%%%%%%%%%%%%%%%%%%%%%%%%%%%%%%%%%%%%%%%%%%%%%%%%%%%%%%%%%%%%%%%%%%%%%%%%%%%%%%%%%%%%%

% SPT PHASE

\paragraph{SPT phase.--} The situation changes for $\mu=0$, for which $\PO$ vanishes in an intermediate region around $J_y/J_x=1$~(inset of Fig.~\ref{fig:pd_cuts_m0}~(b)), 
marking the onset of an intermediate SPT phase. The SPT phase is characterized by the vanishing of all local order parameters that characterize the VA and V0 phases. 
The local density imbalance between the sub-lattices $A$ and $B$
$\Delta n_c = \sum_{i} (-1)^i \langle |\Psi_{i,0}^\dag \Psi_{i,0} - \Psi_{i,1}^\dag \Psi_{i,1}|\rangle$ and $S_R$ are non-zero in the VA phase and zero in the V0 one. Note that the fact 
that $\Delta n_c\neq 0$ in the VA phase implies a spontaneous symmetry breaking of the sub-lattice inversion symmetry.
In contrast,  $S_L\neq 0$ in the V0 phase and zero in the VA one. In the SPT $S_L=S_R=\Delta n_c=0$~(Fig.~\ref{fig:pd_cuts_m0}~(b)).

As in the spin-$1$ Heisenberg model, the SPT phase is protected by a $\mathbb{Z}_2\times \mathbb{Z}_2$ symmetry given by the 
combined set of two orthogonal rotations~\cite{Pollmann2010, Pollmann2012, Pollmann2012A}. 
We choose two transformations that leave $H_{\mathrm{QLL}}$ invariant: ~(${\mathcal{C}}$) particle-hole inversion, $\Psi_{i,j}\leftrightarrow\Psi_{i,j}^\dag$, at all sites 
accompanied by a spin rotation $\sigma^x$ in all bonds, with $\sigma^{x,y,z}$ the Pauli matrices; and~(${\mathcal{R}}$) $\Psi_{i,j}\to -\Psi_{i,j}$ 
for $(i,j)\in A$, and a rotation $\sigma^z$ in all bonds.
Using these two transformations, and following Ref.~\cite{Pollmann2012} we obtain 
from an infinite matrix-product state representation of the ground state
a generalized topological order parameter $\mathcal{O}_T$~\cite{SMnote}.  
In the inset of Fig.~\ref{fig:pd_cuts_m0}~(c) we show $\mathcal{O}_T$ as function of $J_y/J_x$ for $\mu=0$. Whereas  $\mathcal{O}_T=0$ for the V0 and VA phases, 
$\mathcal{O}_T=-1$ in the SPT phase, 
confirming the topologically non-trivial character of the phase.
Contrary to the large-$D$ phase of spin-$1$ chains the VA phase does not display $\mathcal{O}_T=+1$ due to the mentioned spontaneous symmetry breaking of the sub-lattice symmetry.
The topological character of the SPT phase is further confirmed by the doubly degenerate entanglement spectrum shown in Fig.~\ref{fig:pd_cuts_m0}~(c)~\cite{Pollmann2010,Pollmann2012A}.

%%%%%%%%%%%%%%%%%%%%%%%%%%%%

% FIGURE 3

\begin{figure}[bt]
\centering
\includegraphics[width=1\linewidth]{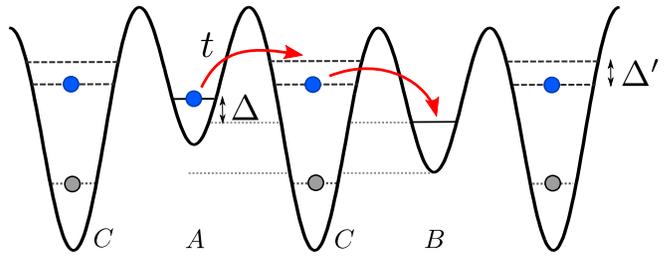}\\
\caption{Sketch of the $s$-$p$ lattice arrangement proposed  for the realization of 1D QLM~(see text).}
\label{fig:scheme}
\end{figure}

%%%%%%%%%%%%%%%%%%%%%%%%%%%%

%%%%%%%%%%%%%%%%%%%%%%%%%%%%%%%%%%%%%%%%%%%%%%%%%%%%%%%%%%%%%%%%%%%%%%%%%%%%%%%%%%%%%%%%%%

% REALIZATION

\paragraph{Realization.--} There have been numerous proposals for the realization of QLM models~\cite{Banerjee2012,Zohar2013,Banerjee2013,Zohar2013b,Tagliacozzo2013,Kasper2017,Hauke2013,Brennen2016}.
Here we introduce a simple scheme~(Fig.~\ref{fig:scheme}), which allows for the dynamical realization of the 1D QLM and QLL, 
based on single-component fermions in an $s$-$p$ lattice formed by deep~(C) and shallow~(A,B) sites similar to that realized in Ref.~\cite{Wirth2011}. 
The lowest state of all C sites, which may be considered as fully pinned, remains filled at any point. 
 We assume two non-degenerate $p$-orbitals, $\alpha=1,2$, in the C sites; the energy splitting $\Delta'$ between both orbitals may be achieved using elliptical sites 
(the third $p$ orbital is assumed to have a much larger energy and can be neglected). Due to the superlattice modulation shallow sites A and B have an energy difference $\Delta$.
The Hamiltonian of the system is 
\begin{eqnarray}
H&=&-t\sum_{k\in {\mathrm A}, k'\in {\mathrm B}}\sum_{\alpha} \left ( \Psi_k^\dag \Phi_{k+1,\alpha} + \Psi_{k'}^\dag \Phi_{k'-1,\alpha} + \mathrm{H.c.} \right )  \nonumber \\ 
&+&  \Delta \sum_{k\in {\mathrm B}} n_k 
+\sum_{k\in {\mathrm C}} \left [ \sum_\alpha E_{\alpha} N_{k,\alpha} + U_{12}N_{k,1}N_{k,2} \right ]
\label{eq:Hphys}
\end{eqnarray}
where $n_k=\Psi_k^\dag \Psi_k$, $N_{k,\alpha}=\Phi_{k,\alpha}^\dag \Phi_{k,\alpha}$,  $t$ denotes the hopping rate between the A~(B) sites and the $p$-orbitals, 
$E_{\alpha}=E_0+\frac{\Delta+(-1)^\alpha \Delta'}{2}+U$, $U$ is the interaction energy between the p-orbitals and the lowest state of the C sites, 
$U_{12}$ is the interaction between p orbitals, and $E_0$ 
is an energy off-set, which can be neglected without loss of generality. We assume  $t\sim |\Delta-\Delta'| \ll U, \Delta, \Delta'$.

The system is initially prepared with a single particle in the $p$ orbitals. 
Due to energy conservation, we may limit ourselves to the manifold in which either $\alpha=1$ or $2$ is occupied at a given C site. 
We may hence introduce  $\tilde S_k^z=\epsilon (N_{k,2}-N_{k,1})/2$, where $\epsilon=1$~($-1$) for C sites at the right~(left) of A sites.
The system then reduces to the 1D QLM with mass $\mu=(\Delta-\Delta')/2$, and $J_x=t^2 U_{12}/U(U+U_{12})$~(for a comparison between Model~\eqref{eq:Hphys} 
and the effective 1D QLM see Ref.~\cite{SMnote}).
An identical scheme may be applied in the $y$ direction to get a 2D QLM, where a possibly different hopping constant results in $J_y$. 
The ladder configuration may be realized by decoupling 
the legs from the rest of the lattice using a sufficiently large energy barriers, as already realized experimentally~\cite{Atala2014}.
We stress that within this setup the actual ground-state is generally not gauge invariant. However, once prepared the gauge-invariant manifold cannot be left within 
second-order processes due to energy conservation. This allows for the dynamical quasi-adiabatic preparation of QLM and QLL states, which we illustrate for the particularly relevant case of the 
SPT phase of the QLL.

%%%%%%%%%%%%%%%%%%%%%%%%%%%%

% FIGURE 4

\begin{figure}[tb]
\centering
\includegraphics[width=0.95\linewidth]{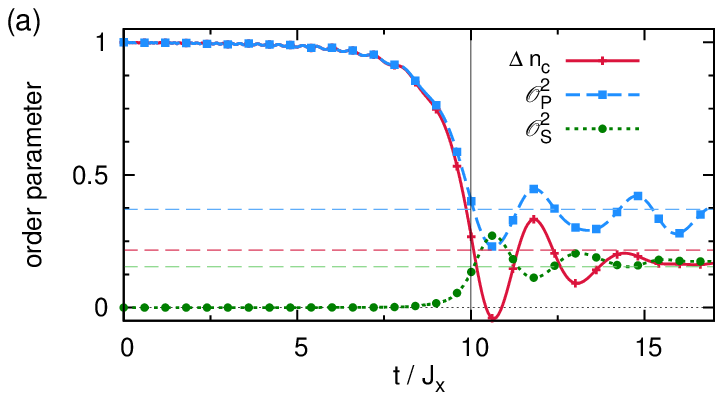}\\
\includegraphics[width=0.95\linewidth]{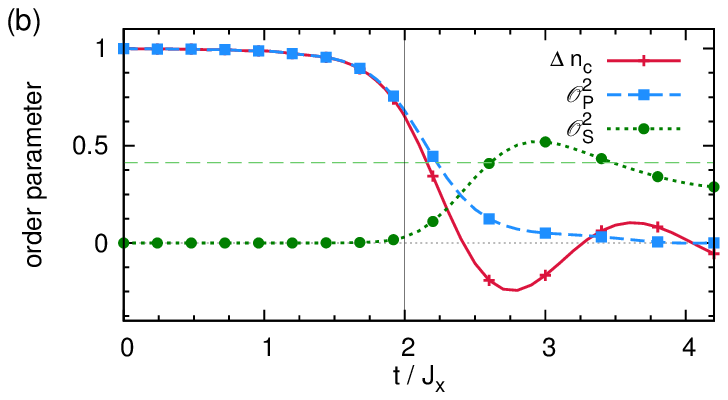}\\
\caption{Quasi-adiabatic preparation. The mass is ramped from $\mu=100 J_x$ down to $\mu=0$ as $\mu\sim (t-t_R)^{1/4}$. 
Figures (a) and (b) show the time evolution of $\PO$, $\SO$, and $\Delta n_c$ for, respectively, 
a 1D QLM  with $J_x t_R=10$, and a QLL with $J_x t_R=2$. The results have been obtained using iTEBD with up to $800$ matrix states.
Dashed horizontal lines indicate the expected values of the order parameters in the ground-state with $\mu=0$. 
Note that in both cases the quasi-adiabatic ramp leads to a finite $\SO$. In contrast, for the 1D QLM $\PO$ and $\Delta n_s$ oscillate around 
the expected finite value, whereas $\PO,\Delta n_c\simeq 0$ in the QLL, as expected for the SPT phase. 
}
\label{fig:ramp_ladder}
\end{figure}

%%%%%%%%%%%%%%%%%%%%%%%%%%%%

The defect free VA phase is a product state that may be prepared by filling all B sites, keeping A sites empty, and filling the deepest and $\alpha=1$ state of C sites. 
Note that the preparation of this initial state fixes the boundary conditions of the QLL~(grey spins in Fig.~\ref{fig:patterns}). 
Starting at large $\mu\gg J_{x,y}$, non-trivial quantum many-body states may be prepared by a quasi-adiabatic decrease of the mass $\mu$.  Note in this sense that 
neither for the 1D nor for the ladder case a phase transition is encountered, and hence $\mu=0$ states may be prepared in a finite time without crossing a quantum critical point. 
In Fig.~\ref{fig:ramp_ladder}(b) we show for the case of $J_x=J_y$ that a short ramping sequence ($t_{R}\sim J_x$) is sufficient to prepare quantum states at $\mu=0$ 
with properties very similar to the SPT state.  Although due to the finite ramp the expectation values oscillate, these values are close to the ground-state expectation~(dashed lines) 
showing $\PO\simeq 0$ but $\SO>0$ as expected for the SPT phase. In contrast, a similar preparation for the 1D QLM~(Fig.~\ref{fig:ramp_ladder}(a)) results, as
expected, in $\PO,\SO\neq 0$.

%%%%%%%%%%%%%%%%%%%%%%%%%%%%%%%%%%%%%%%%%%%%%%%%%%%%%%%%%%%%%%%%%%%%%%%%%%%%%%%%%%%%%%%%%%

% SUMMARY

\paragraph{Summary.--} We have shown that quantum link ladders present an intriguing phase diagram characterized by the appearance of a symmetry protected topological 
phase, which is revealed by a non-local spin string order along each of the ladder legs. We have discussed a simplified dynamical realization that permits the (quasi)~adiabatic creation of 
the states of the quantum link models, and in particular the topological phase. Our results open intriguing questions about the possibility to observe similar phases and edge string-order
in finite two-dimensional quantum-link lattices.

%%%%%%%%%%%%%%%%%%%%%%%%%%%%%%%%%%%%%%%%%%%%%%%%%%%%%%%%%%%%%%%%%%%%%%%%%%%%%%%%%%%%%%%%%%

% ACKNOWLEDGEMENTS

\begin{acknowledgments}
We acknowledge support of the German Research Foundation DFG (projects RTG 1729 and no. SA 1031/10-1).
Simulations were carried out on the cluster system at the Leibniz University of Hannover, Germany.
\end{acknowledgments}

%%%%%%%%%%%%%%%%%%%%%%%%%%%%%%%%%%%%%%%%%%%%%%%%%%%%%%%%%%%%%%%%%%%%%%%%%%%%%%%%%%%%%%%%%%

\bibliography{references}

\begin{thebibliography}{53}
\bibitem [{oth()}]{otherstatesnote}%
  \BibitemOpen
  \href@noop {} {}\bibinfo {note} {{\rm States $ \left | \begin{array}{c} \pm
  \\ \mp \end{array} \right \rangle$ exist as well, and have a similar energy
  up to second-order as $|\phi_{\pm 1}\rangle$, but higher-order ring-exchange
  terms favor $|\phi_{\pm 1}\rangle$, and hence they are neglected in the
  discussion.}}\BibitemShut {Stop}%
\end{thebibliography}

\end{document}

\title{Supplementary Material to ''Hidden order and symmetry protected topological states in quantum link ladders''}

\author{L. Cardarelli}
\author{S. Greschner}
\author{L. Santos}
\affiliation{Institut f\"ur Theoretische Physik, Leibniz Universit\"at Hannover, 30167~Hannover, Germany} 

%%%%%%%%%%%%%%%%%%%%%%%%%%%%%%%%%%%%%%%%%%%%%%%%%%%%%%%%%%%%%%%%%%%%%%%%%%%%%%%%%%%%%%%%

\begin{abstract}
In this Supplementary Material we discuss additional details concerning the characterization of the ground-state phases and dynamics of the 1D QLM and the QLL discussed in the main text.
\end{abstract}

%%%%%%%%%%%%%%%%%%%%%%%%%%%%%%%%%%%%%%%%%%%%%%%%%%%%%%%%%%%%%%%%%%%%%%%%%%%%%%%%%%%%%%%%

\date{\today}

\maketitle

%%%%%%%%%%%%%%%%%%%%%%%%%%%%%%%%%%%%%%%%%%%%%%%%%%%%%%%%%%%%%%%%%%%%%%%%%%%%%%%%%%%%%%%%

% 1D QLM

\section{1D QLM}

%%%%%%%%%%%%%%%%%%%%%%%%%%%%%%%%%%%%%%%%%%

\subsection{Ground-state properties}

\begin{figure}[t]
\centering
\includegraphics[width=0.8\linewidth]{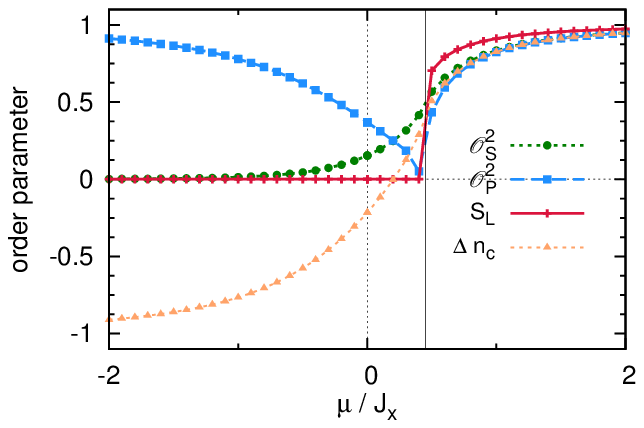}
\caption{Ground-state order parameters for the 1D QLM as function of $\mu/J_x$. The results were obtained using iTEBD with $100$ states. The solid vertical line marks the phase transition 
from the Z~($\mu< \mu_c\sim 0.45 J_x$) to the NZ phase~($\mu> \mu_c$).}
\label{fig:cuts_chain_order}
\end{figure}

Figure~\ref{fig:cuts_chain_order} depicts $\PO$ and $\SO$ evaluated using infinite time evolving block decimation (iTEBD) simulations~\cite{Vidal2007}. 
At $\mu/J_x\simeq 0.45$ a phase transition separates the Z and the NZ phase. 
The NZ phase exhibits at any $\mu$ a finite magnetization $S_L=\frac{1}{L}\sum_i \langle |S^z_{i,i+1}|\rangle $. In both phases 
$\PO, \SO\neq 0$ due to the explicitly broken chiral symmetry. For a detailed discussion of the 1D QLM in a similar context see Ref.~\cite{Rico2014}.

%%%%%%%%%%%%%%%%%%%%%%%%%%%%%%%%%%%%%%%%%%

%%%%%%%%%%%%%%%%%%%%%%%%%%%%%%%%%%%%%%%%%%

\subsection{Comparison with the $s$-$p$ model}

\begin{figure}[t]
\centering
\includegraphics[width=0.8\linewidth]{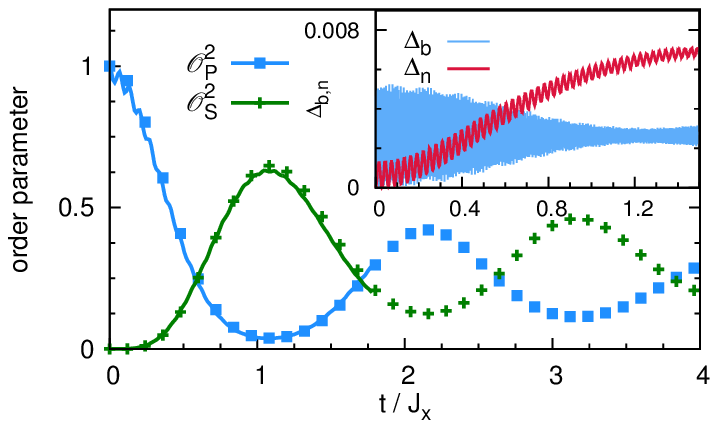}
\caption{Emergence of $\SO$ after a sudden quench. We compare the time evolution of $\SO$ and $\PO$  for  the effective 1D QLM and for Model~(4) of the main text with $\Delta=4\tilde{J}$ and $U=U_{12}=40\tilde{J}$. For the full model only the total average particle number per unit cell is fixed. The inset shows the deviation of Model~(4)  $\Delta_n$ and $\Delta_b$ (see text). The iTEBD simulations are terminated after a limit of $800$ matrix 
states is reached.}
\label{fig:quench_chain_full_eff}
\end{figure}

In the main text we have discussed the quasi adiabatic preparation of the $\mu\to 0$ states in the 1D QLM (and the QLL) by means of a ramping of the mass term during a finite time.
In Fig.~\ref{fig:quench_chain_full_eff} we analyze the case of a sudden quench of $\mu$ for the 1D QLM and the comparison to the time evolution of $s$-$p$ 
Model~(4) of the main text. Interestingly, already the sudden quench situation exhibits a finite $\SO, \PO>0$.
Both the time evolution of the effective QLM and of Model~(4) of the main text agree accurately~(note that due to numerical limitations we only follow the time evolution of the $s$-$p$ model during a shorter time). 
In order to quantify the accuracy with Model (4) realizes an effective QLM we study the deviation from Gauss' law~\cite{Banerjee2012}. The parameter
\begin{align}
\Delta_n = \frac{1}{L} \sum_{k\in A,B} | N_{k-1,1} + n_k + N_{k+1,2} - 2 |
\end{align}
measures the deviation of the occupation of particles on neighboring sites~($\Delta_n=0$ for a perfect QLM realization). We furthermore analyze whether the $p$-orbitals form a spin $1/2$, i.e. if precisely one $p$-orbital is occupied. To this aim we introduce 
\begin{align}
\Delta_b = \frac{1}{L} \sum_{k\in C} | N_{k,1} + N_{k,2} - 1 | \,,
\end{align}
which is zero for a perfect QLM realization.  For the parameters of Fig.~\ref{fig:quench_chain_full_eff} both $\Delta_n$ and $\Delta_b<10^{-2}$~(inset of Fig.~\ref{fig:quench_chain_full_eff}).
Hence, with the system initialized as a gauge invariant product state, Gauss' law can be fulfilled during a sufficiently long time evolution that 
allows for the observation of nontrivial $\PO$ and $\SO$ correlations.

%%%%%%%%%%%%%%%%%%%%%%%%%%%%%%%%%%%%%%%%%%

%%%%%%%%%%%%%%%%%%%%%%%%%%%%

% FIGURE 1

%%%%%%%%%%%%%%%%%%%%%%%%%%%%

%%%%%%%%%%%%%%%%%%%%%%%%%%%%

% FIGURE 2

%%%%%%%%%%%%%%%%%%%%%%%%%%%%

%%%%%%%%%%%%%%%%%%%%%%%%%%%%%%%%%%%%%%%%%%%%%%%%%%%%%%%%%%%%%%%%%%%%%%%%%%%%%%%%%%%%%%%%%%

% QLL

\section{QLL}

%%%%%%%%%%%%%%%%%%%%%%%%%%%%%%%%%%%%%%%%%%

\subsection{Ground-state phases and phase transitions}

\begin{figure}[t]
\centering
\includegraphics[width=0.8\linewidth]{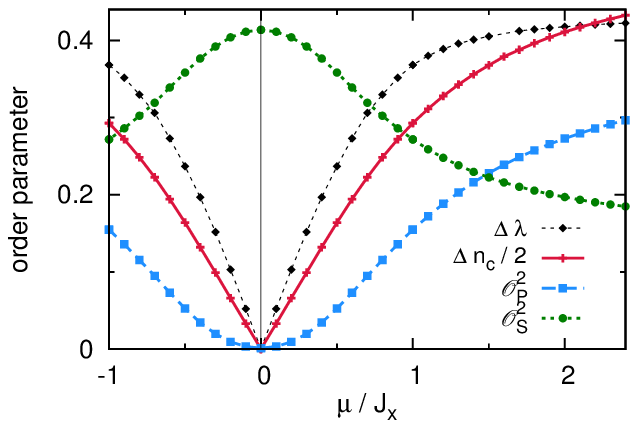}
\caption{Cut through phase diagram Fig.~2~(a) of the main text for $J_y=J_x$ as function of the mass $\mu$ (iTEBD simulations with $100$ states). Only for $\mu=0$ a SPT phase is realized and $\PO$ vanishes while 
$\SO$ remains finite.}
\label{fig:cuts_fm}
\end{figure}

Figure~\ref{fig:cuts_fm} shows our iTEBD results for the QLL with $J_x=J_y$ as a function of $\mu$. 
As mentioned in the main text, the mirror symmetry of the ladder results in a symmetry $\mu\leftrightarrow -\mu$ for the QLL.
$\PO$~(as well as other local order parameters) immediately increases when $|\mu|>0$, whereas $\SO$ remains finite and only vanishes in the very limit of $|\mu|\to\infty$. 
Note that  due to the broken inversion-, particle-hole-, and sublattice-symmetry, the SPT phase adiabatically connects to the VA phase.

A necessary property of SPT phases is the double degeneracy in the entanglement spectrum~\cite{Pollmann2010,Pollmann2012A}. 
In Fig.~\ref{fig:cuts_fm} we depict as well the entanglement gap $\Delta \lambda = \sum_i (-1)^i \lambda_i$, where 
$\lambda_i$ is the ordered sequence of Schmidt eigenvalues. Only for $\mu=0$ this values vanishes.

%%%%%%%%%%%%%%%%%%%%%%%%%%%%

% FIGURE 3

%%%%%%%%%%%%%%%%%%%%%%%%%%%%

%%%%%%%%%%%%%%%%%%%%%%%%%%%%

% FIGURE 4

\begin{figure}[t]
\centering
\includegraphics[width=0.8\linewidth]{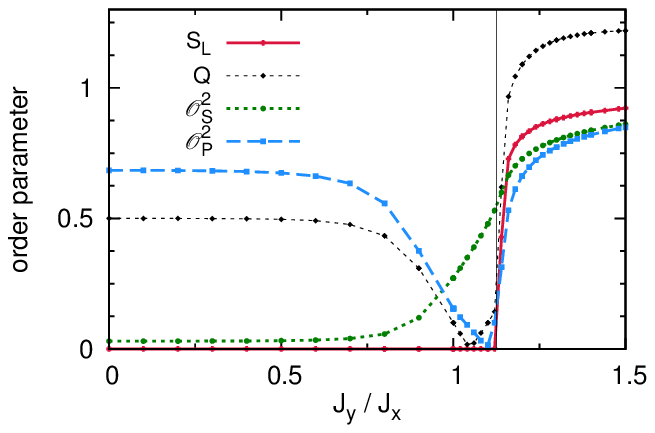}
\caption{Cut for $\mu=J_x$ of the phase diagram Fig.~2~(a) of the main text as function of $J_y/J_x$~(iTEBD simulations with $100$ states).}
\label{fig:cuts_m1_order}
\end{figure}

%%%%%%%%%%%%%%%%%%%%%%%%%%%%

%%%%%%%%%%%%%%%%%%%%%%%%%%%%

% FIGURE 5

\begin{figure}[b]
\centering
\includegraphics[width=0.8\linewidth]{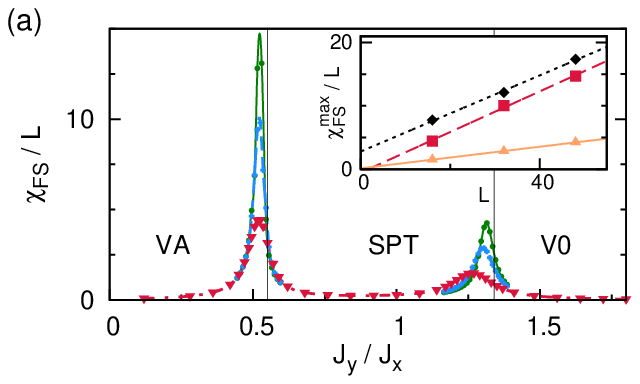}
\includegraphics[width=0.8\linewidth]{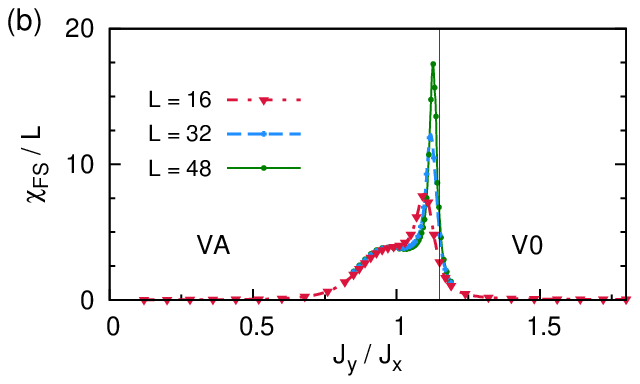}
\caption{Scaling of the fidelity susceptibility $\chi_{FS} /L $ for the QLL as function of $J_y/J_x$, for (a) $\mu/J_x=0$ and (b) $\mu/J_x=1$. The results are obtained from DMRG-simulations keeping up to 
$800$ states. The inset of (a) shows a linear scaling of the peak of the $\chi_{FS} /L$-curve with the number of rungs $L$, proving the Ising character of the quantum phase transitions 
(from bottom to top) between SPT to V0~($\mu=0$), VA to SPT~($\mu=0$) and VA to V0~($\mu=J_x$).}
\label{fig:cuts_m0_1_fs}
\end{figure}

%%%%%%%%%%%%%%%%%%%%%%%%%%%%

Figure~\ref{fig:cuts_m1_order} depicts the order parameters for a cut through the phase diagram of Fig.~2~(a) of the main text. We only observe one VA-to-V0 phase transition for $J_y\sim J_x$, 
marked by the abrupt growth of the leg magnetization $S_L$. In order to further characterize the transitions we analyze by means of density-matrix-renormalization-group~(DMRG)~\cite{White1992, Schollwoeck2011} calculations 
the fidelity susceptibility 
\begin{align}
\chi_{FS}(U) = \lim_{\delta U\to 0} \frac{-2 \ln |\langle \Psi_0(U) | \Psi_0(U + \delta U) \rangle| }{(\delta U)^2} \,,
\end{align}
with $|\Psi_0\rangle$ being the ground-state wave function. Marked peaks reveal the presence of two phase transitions for $\mu=0$, and a single one for $\mu\neq 0$. 
The scaling of the peak ${\rm max} \chi_{FS}(\phi)$ with the system size is consistent with a second-order Ising-like character for all transitions.

%%%%%%%%%%%%%%%%%%%%%%%%%%%%%%%%%%%%%%%%%%

%%%%%%%%%%%%%%%%%%%%%%%%%%%%%%%%%%%%%%%%%%

\subsection{Topological order parameter}

We obtain the generalized topological order parameter following the procedure of Ref.~\cite{Pollmann2012}.  From a canonical infinite matrix-product state~(IMPS) representation of the ground state, 
$|\Psi\rangle = \sum_\sigma \Lambda \Gamma_\sigma | \sigma \rangle$, we evaluate the eigenvalues $\eta_{\mathcal{C,R}}$ of the generalized transfer matrices 
$T_{\mathcal{C,R}} = \sum \kappa_{\sigma\sigma'}^{\mathcal{C,R}} \Gamma_\sigma \Gamma_{\sigma'} $, 
with $\kappa_{\mathcal{C,R}}$ the unitary matrices of the symmetries ${\mathcal{C}}$ and ${\mathcal{R}}$ . From the corresponding eigenstates we obtain the projective matrix representation of the symmetries 
$\mathcal{U}_\mathcal{R}$, $\mathcal{U}_\mathcal{C}$. The generalized topological order parameter is given by
\begin{align}
\mathcal{O}_T = \begin{cases}
0&\text{ if } |\eta_\mathcal{C}|<1 \text{ or } |\eta_\mathcal{R}|<1\\
\frac{1}{\chi} \tr\left(\mathcal{U}_\mathcal{C} \mathcal{U}_\mathcal{R} \mathcal{U}_\mathcal{C}^\dagger \mathcal{U}_\mathcal{R}^\dagger \right) &\text{ if } |\eta_\mathcal{C}|=|\eta_\mathcal{R}|=1  
\end{cases}
\end{align}

%%%%%%%%%%%%%%%%%%%%%%%%%%%%%%%%%%%%%%%%%%%%%%%%%%%%%%%%%%%%%%%%%%%%%%%%%%%%%%%%%%%%%%%%%%

\bibliography{references}